# NMR Probe of Metallic States in Nanoscale Topological Insulators

*Dimitrios Koumoulis[†], Thomas C. Chasapis*, Robert E. Taylor[†], Michael P. Lake[†], Danny King[†], Nanette N. Jarenwattananon[†], Gregory A. Fiete[††], Mercouri G. Kanatzidis\* and Louis-S. Bouchard[†‡§]*

[†]Department of Chemistry and Biochemistry, University of California, Los Angeles, Los Angeles, CA 90095, USA

[††] Department of Physics, University of Texas at Austin, Austin, TX 78712 USA

[§]California NanoSystems Institute, UCLA, Los Angeles, CA 90095, USA

*Department of Chemistry, Northwestern University, 2145 Sheridan Rd, Evanston, IL 60208, USA

[‡] Corresponding author's email address:

bouchard@chem.ucla.edu

Last revised: October 30, 2012

**Abstract**
A $^{125}$Te NMR study of bismuth telluride nanoparticles as function of particle size revealed that the spin-lattice relaxation is enhanced below 33 nm, accompanied by a transition of NMR spectra from single to bimodal regime. The satellite peak features a negative Knight shift and higher relaxivity, consistent with core polarization from p-band carriers. Whereas nanocrystals follow a Korringa law in the range 140-420K, micrometer particles do so only below 200K. The results reveal increased metallicity of these nanoscale topological insulators in the limit of higher surface-to-volume ratios.

**Introduction**

Physical particles, when small enough, are known to display properties which differ from the bulk material due to both surface effects and quantum size effects [1-2]. Examples of physical properties affected by particle size include melting points, reactivity, magnetism, and the fluorescence of quantum dots [1, 3]. For metals and semiconductors, the density of states undergoes substantial changes as the particle size is less than the de Broglie wavelength, because of altered electronic properties [1, 4]. Bismuth telluride ($Bi_2Te_3$) is a narrow-gap semiconductor known for its thermoelectric properties [5-7] and more recently as a topological insulator (TI). TI materials are characterized by insulating band gaps in the bulk and a gapless surface state which is a Dirac-like cone as a result of spin-orbit coupling and time-reversal symmetry [8-10].

The characterization of TI states requires sensitive techniques to probe metallicity of the surface. To date, electrical transport measurements, *s*canning *t*unneling *m*icroscopy (STM) and *a*ngle-*r*esolved *p*hoto-*e*mission *s*pectroscopy (ARPES) have been the main workhorses in the field. These techniques work best with high-quality thin films (<20 nm) or large single crystals and low temperatures (<30 K). On the other hand, there is a need for characterizing materials at room temperature and materials of suboptimal quality. *N*uclear *m*agnetic *r*esonance (NMR) spectroscopy is known to be a highly sensitive probe of the electronic wavefunction in semiconductors, but has not yet been investigated as a potential tool for probing the surface states of TI. Advantages of NMR may include operation at higher temperatures and the ability to probe the properties of lower quality or even amorphous materials. This could enable the investigation of entirely new classes of TI materials, such as "granular" varieties or those with a large number of bulk defects with topologically protected gapless modes [11] which are not suitable for study by conventional techniques. Another potential advantage of NMR is its ability to perform sensitive magnetometry non-invasively, which we anticipate could play a useful role in the study and manipulation of spin states in TIs. This is the first NMR study to present evidence of a link between nuclear magnetism and metallic states in TIs.

Previously, we have carried out a $^{125}$Te NMR study [12] of $Bi_2Te_3$ powders ground by mortar and pestle (m&p) and found that a Korringa law developed below 200 K. The m&p samples reflect properties of the bulk because of the fairly large (μm) average grain size. The development of the Korringa law for $Bi_2Te_3$ m&p powders below 200 K is shown by the blue curve in Figure 1. Above 320 K, the relaxation data showed thermally-activated relaxation arising from interaction with charge carriers with an activation energy of 8.44 kJ/mol (87 meV). This activation energy may correspond to excitations of electrons from impurity states into the conduction band or excitations across the bulk bandgap. The latter possibility cannot be ruled out given the high uncertainty in fitting activation energies over the limited temperature range 140-420 K. The origin of the Korringa process in these narrow-gap insulators, however, warrants further investigations of the spin interactions, especially in the regime where surface effects dominate. We find increased metallicity in nanoscale TIs in the limit of higher surface-to-volume ratios, up to room temperatures.

$^{125}$Te NMR data were acquired with a Bruker DSX-300 spectrometer operating at 94.79 MHz using a standard Bruker X-nucleus wideline probe with a 5-mm solenoid coil. The $^{125}$Te $\pi/2$ pulse width in the wideline probe was 4 μs. $Bi_2Te_3$ ingots were reduced to powders of various sizes by m&p and ball-milling under inert atmosphere. Particle sizes reported are mean values determined from powder x-ray diffraction (PXRD) data. PXRD was also used to confirm crystallinity of the samples after ball milling. Spectral data were acquired using a spin-echo sequence [$\pi/2)_x - \tau - \pi)_y$ - *acquire*]. The echo delay, τ, was set to 20 μs. The $^{125}$Te chemical

shift scale was calibrated using the unified $\Xi$ scale [13], relating the nuclear shift to the $^1$H resonance of dilute tetramethylsilane in CDCl$_3$ at a frequency of 300.13 MHz. In order to acquire the full $^{125}$Te NMR spectrum of ball-milled Bi$_2$Te$_3$, we used the *v*ariable *o*ffset *c*umulative *s*pectra (VOCS) technique [14]. $T_1$ data were acquired with the saturation-recovery technique [15] and fitted to the Kohlrausch (stretched exponential) function:

$$M(t) = M_0[1 - \exp(-(t/T_1)^\beta)] \qquad (1)$$

In the ball milled samples, $\beta = 0.5$ provided a good fit, whereas for m&p sample, $\beta = 1$ provided a better fit.

The first results for ball-milled Bi$_2$Te$_3$ are shown in Figure 1, where the red trace is the Korringa value, $T_1 \cdot T$, as function of inverse temperature, $1/T$, for nanocrystals of mean diameter ~19 nm. The Korringa values, $T_1 \cdot T$, were 24 $s \cdot K$ for the m&p sample (below 200 K) and 11 $s \cdot K$ for the nanocrystals (across all temperatures, 140-420 K). These values are not unusual for semiconductors and tend to reflect the fact that $T_1 \cdot T$ is inversely proportional to the square of the electronic density of states (DOS) at the Fermi level, $N^2(E_F)$ [16-21]. A lower Korringa value means higher carrier density and a more metallic state. These data indicate that 19 nm particles are more metallic than the m&p sample. In Figure 1 – inset we plot $1/(T_1 \cdot T^{1/2})$ versus $1/T$. A straight line is indicative of a metallic sample, where the quantity $1/(T_1 \cdot T^{1/2})$ is proportional to the carrier density [17-22]. The inset graph shows that the nanocrystals have higher carrier density than the m&p samples across all temperatures investigated.

Further insights can be gained into the relaxation mechanism by measuring $1/T_1$ versus particle size (Fig. 2a). A range of sizes (19 nm – 1 μm) was obtained by ball milling m&p samples for different amounts of time – from 0 to 732 min (Fig. 2b). A *s*canning *e*lectron *m*icroscopy (SEM) image for particles with 55 nm average size revealed a polydisperse distribution [Fig. 2 (d)]. The relaxation rate is seen to sharply increase three-fold when the particle size drops below 33 nm. If the increase in relaxation rate were simply due to increased defects from crystallite damage, $1/T_1$ would be expected to progressively increase with decreasing particle sizes. Instead, changes occur abruptly, as evidenced by the sigmoidal shape in the graph of $1/T_1$ vs particle size [Fig. 2 (a)]. This suggests that the change in the spin-lattice relaxation rate is probing a particle size (electronic) effect rather than progressively increased damage. The possibility of damage to the bulk, however, must be ruled out, according to the PXRD spectra for m&p and ball milled (732 min.) samples [Fig. 2 (b) inset], which indicate that the nanoparticles possess an identical crystal structure as the bulk. This shows that the observed changes in $T_1$ are likely due to surface effects.

We recall that ideal 3D TIs are insulators in the bulk and conductive on their surfaces. The thickness of the metallic surface layer has been measured in transport studies to be approximately 3 quintuple layers (QL) [23-25]. At the transition point (33 nm), the volume fraction of surface states (3 QL thick) of these nanocrystals is nearly 50%. For 19 nm particles, the volume fraction of surface states reaches almost 70%. Thus, the abrupt change occurs when particles are in the limit of large surface-to-volume ratios. It is interesting to note that the data of Figure 2 was acquired at room temperature, meaning that these metallic effects can manifest themselves at ambient conditions.

A naïve core-shell model, such as illustrated in [Fig. 2 (a) inset], where $1/T_1$ is the sum of bulk ($1/T_{1,\text{bulk}}$) and surface ($1/T_{1,\text{surf}}$) spin-lattice relaxation rates ($1/T_{1,\text{bulk}} < 1/T_{1,\text{surf}}$) each weighted by their respective volume fractions, does not fully describe the data of Fig. 2 (a). In particular, it fails to reproduce the sudden change in relaxation rate and the apparent "plateau" below 33 nm. A competing effect is needed to modulate the dependence of the $1/T_1$ rate on particle size. Such a competing effect could arise from spin diffusion or a slight drop in carrier concentration upon ball milling. Nuclear-spin diffusion [22,26] between the bulk and surface, which are expected to be important in the limit of small particle sizes, would have the effect of mixing the relaxation rates of bulk and surface spins, leading to a slight de-emphasis of each phase. A second possible competing effect could be a slight decrease in carrier density when going to small sizes, as material defects tend to be pushed toward the surface. The particle size distribution and shape [clearly non-spherical, e.g., Fig. 2 (d)] enter the picture indirectly, by modulating the effects of nuclear-spin diffusion. What is clear from these results, however, is that as the mean particle size is reduced, an increasingly metallic character is observed by NMR.

Further signs of metallicity are found by analyzing the $^{125}$Te NMR spectrum as function of particle size [Fig. 2 (c)]. The lineshape for the m&p sample, while slightly asymmetrical, is essentially unimodal. Its origin, which has been described previously [12], includes contributions from shielding anisotropy [27-29] from the low symmetry of the rhombohedral space group [8], $D_{3d}^5$ ($R\bar{3}m$), as well as possible lattice defects [30-31]. The nanocrystals, however, are distinguished by the presence of a satellite peak with a negative Knight shift. Negative Knight shifts in p-type semiconductors are classical cases of core polarization from p-band carriers [31, 32-36]. The satellite peak amplitude increases with decreasing particle size [Fig. 2 (c)]. Moreover, the center frequency of the satellite peak shifts toward more negative frequencies with decreasing particle size (Fig. 3 inset), suggesting an increasingly more metallic state. The central peak frequency, however, remains mostly independent of particle size (Fig. 3 inset). The bifurcation, although small, does exist even at 55 nm. This could be an effect from the "tail" of the particle size distribution. We also note that for an average particle size of 55 nm, surface layers still represent significant volume fraction (~30%, assuming 3 QL-thick surface layers in a core-shell geometry), which could account for the residual Knight shift of the satellite peak at that size. In any case, this feature may deserve further study, perhaps with first-principles methods.

We have fitted $T_1$ from a saturation-recovery experiment at room temperature for the central vs satellite peaks separately and found significantly different $T_1$ values for each peak (Fig. 4). Namely, the shoulder peak relaxes 2.3 times faster than the central peak. Since $\frac{1}{T_1 \cdot \sqrt{T}} \propto N$, we estimate that the shoulder peak has 2.4 times higher carrier density compared to the bulk (central peak). This result is consistent with the premise that $T_1$ dispersion across frequencies is governed by interactions with the conduction carriers. The increased relaxation rate of the satellite peak indicates a higher carrier concentration and hence, more metallic state. Given this evidence, we infer that the central peak corresponds to bulk states whereas the satellite peak corresponds to more metallic states.

A word about the possible contribution of defects is in order. The NMR spin-lattice relaxation rate, $1/T_1$, in semiconductors is proportional to carrier density [37]. The presence of defects, impurities or dopants alters the electronic properties[33,34]. Previous studies on PbTe samples ball milled for 60 min [38-39] show that crystallinity is preserved, but carrier concentration decreases. The decrease in carrier concentration in nanocrystalline samples is referred to as the "self-cleaning" process of the bulk [40-43] whereby defects are pushed toward the surface. The nanoparticles possess cores with fewer defects than the corresponding bulk material. Because of self-cleaning in nanoparticles, ball-milling would be expected to increase $T_1$ and decrease linewidth. In our case, however, the opposite effect is observed: $T_1$ decreases, and the spectrum becomes wider (includes a satellite peak with negative Knight shift and shorter $T_1$). In light of the observed abrupt changes in Knight shift and $T_1$ effects, as well as PXRD results which confirm structural integrity of the bulk [Fig. 2 (b) inset], the contribution from defects, while it cannot completely be excluded, must also be regarded as a minor effect.

In Figure 3, a plot of the main peak width as a function of particle size, again at ambient temperature, shows that the main peak broadens with decreasing particle size and the transition is sharp near 33 nm, [The two peaks can be seen in spectra of Figure 2 (c)]. This broadening could be due to several possible contributions: (1) increased surface-to-volume ratio exposing metallic states whose conduction electrons cause homogeneous broadening by spin scattering, (2) increased particle size distribution (variance) and associated Knight shift from the increased variance. Increased broadening with decreasing particle due to a Knight shift would provide additional evidence that the observed transition below 33 nm is associated with metallic behavior.

In this study, the $^{125}$Te spin-lattice relaxation rates of nanocrystalline $Bi_2Te_3$ was found to be fundamentally different from micrometer size powders: a Korringa process is observed across all temperatures whereas the $1/T_1$ rate is seen to triple below particle diameters of 33 nm. Moreover, a negative Knight shift is observed in nanocrystals as function of particle size, which suggests increasing metallic behavior in the limit of large surface-to-volume ratios. The PXRD results [Fig. 2 (b) inset] do not provide evidence of any damage to the bulk, suggesting that the observed metallicity is likely primarily a surface effect. Also, increased damage to the bulk would have been expected to show a gradual change in $T_1$ as opposed to the abrupt change observed below 33 nm. This behavior suggests that the change in spin-lattice relaxation behavior results from surface effects which tend to dominate the NMR signal in the nanocrystalline regime, where the surface-to-volume ratio is high. The metallic behavior is observed even at room temperature, suggesting a potentially new way to probe metallicity in these materials. Preliminary results in different TI and non-TI materials (Table 1) also suggest that this method may be applicable to other materials. In Table 1, the TI materials ($Bi_2Te_3$, and $Bi_2Se_3$) all show a substantial drop in $T_1$ upon ball milling – suggesting higher metallicity – whereas the non-TIs (PbTe and ZnTe) instead shows an increase in $T_1$. For all samples in Table 1, PXRD results (not shown) confirmed that the nanocrystals preserved their structural integrity after ball milling. Generally, when semiconductor nanoparticle size decreases dopants are pushed to the surface and expelled reducing the carrier concentration and consequently the metallicity. In topological insulators however, surface metallicity is protected by time-reversal symmetry and spin-orbit coupling and should persist as size decreases.

While the NMR technique does not directly measure the Dirac cone as ARPES would, it does provide results consistent with metallic surface states and could find applications for topological materials beyond TIs. The material studied herein, $Bi_2Te_3$, has a Dirac point buried in the bulk valence band and a Fermi level which rises well into the conduction band [8,44,45]. In general, $T_1$ is expected to be shorter with a higher density of states, which would happen if the Fermi energy is in an energy range of both surface and bulk bands; this is consistent with results in Table I. The key feature of the NMR probe is that by studying TI systems on the nanoscale the problem of core metallicity interfering with the surface state in many cases may be bypassed. Nanoscale TI samples therefore have the surface contribution significantly enhanced. In this case, NMR will be a powerful characterization tool in observing and studying the TI surface states.

**Acknowledgements**

The work at UCLA and Northwestern was supported by DARPA (MTO) MESO; at UCLA by AFOSR/DARPA (DSO) QuASAR (magnetometry portion – Figure 3 inset); at UT by ARO Grant W911NF-09-1-0527 and NSF Grant DMR-0955778. L.-S. B. acknowledges useful discussions with S.D. Mahanti and thanks Yavuz Ertas for assistance with SEM.

**Tables**

**Table 1. Comparison of $T_1$ relaxation times for various binary TI and non-TI materials.**
For the TI materials studied (Bi$_2$Se$_3$ and Bi$_2$Te$_3$), the $T_1$ decreases upon ball milling whereas for the non-TIs (PbTe and ZnTe), it increases. All ball milling times resulted in average particle sizes in the range 30-50 nm.

| Material | $T_1$ m&p | $T_1$ ball milled | milling time |
|---|---|---|---|
| Bi$_2$Te$_3$ | 133 ms | 38 ms | 30 min. |
| Bi$_2$Se$_3$ | 9.6 s | 1.92 s | 180 min. |
| ZnTe | 173.8 s | 376.3 s | 180 min. |
| PbTe | 2.6 s | 12.7 s | 60 min. |

**Figures and Figure Captions**

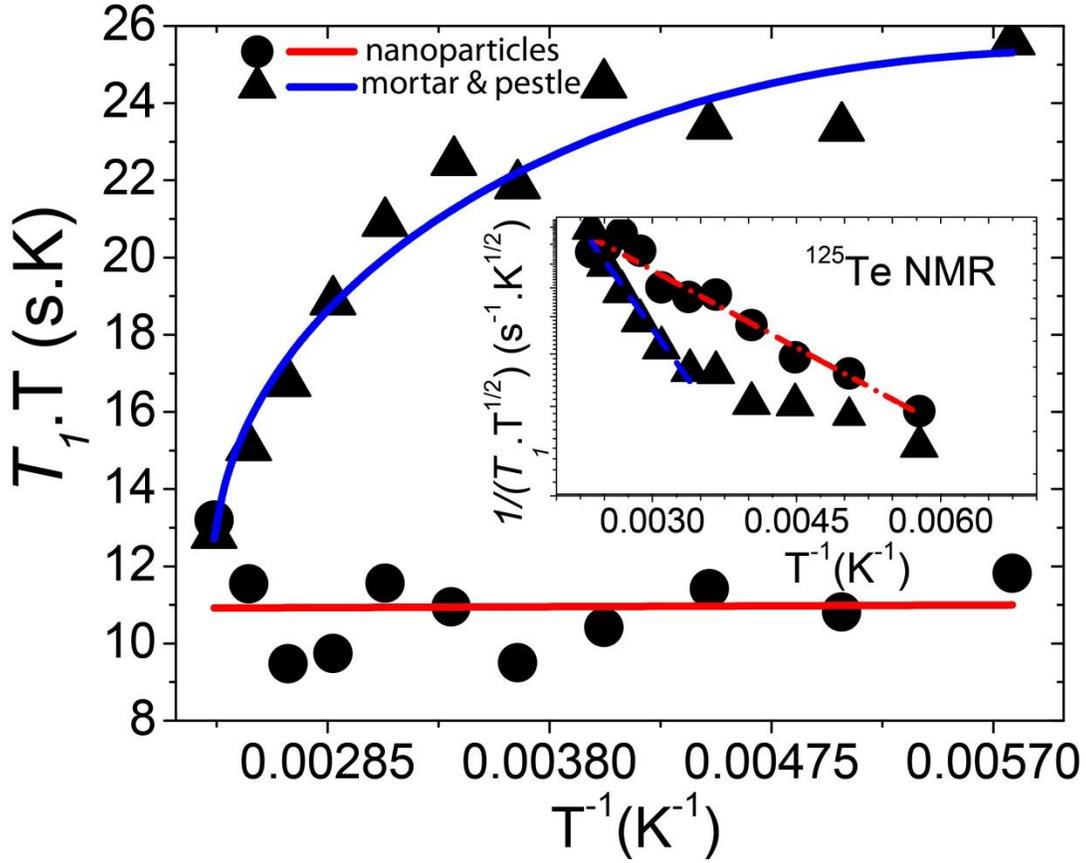

FIG. 1 (color online) Comparison of the Korringa product of the $^{125}$Te spin-lattice relaxation time with the temperature as function of the inverse temperature for 19-nm nanoparticles of $Bi_2Te_3$ and mortar & pestle (m&p) sample. Inset shows a log-log plot for $1/(T_1 \cdot T^{1/2})$ vs inverse temperature in the case of nanosized as well as m&p samples. The quantity $1/(T_1 \cdot T^{1/2})$ is proportional to carrier density.

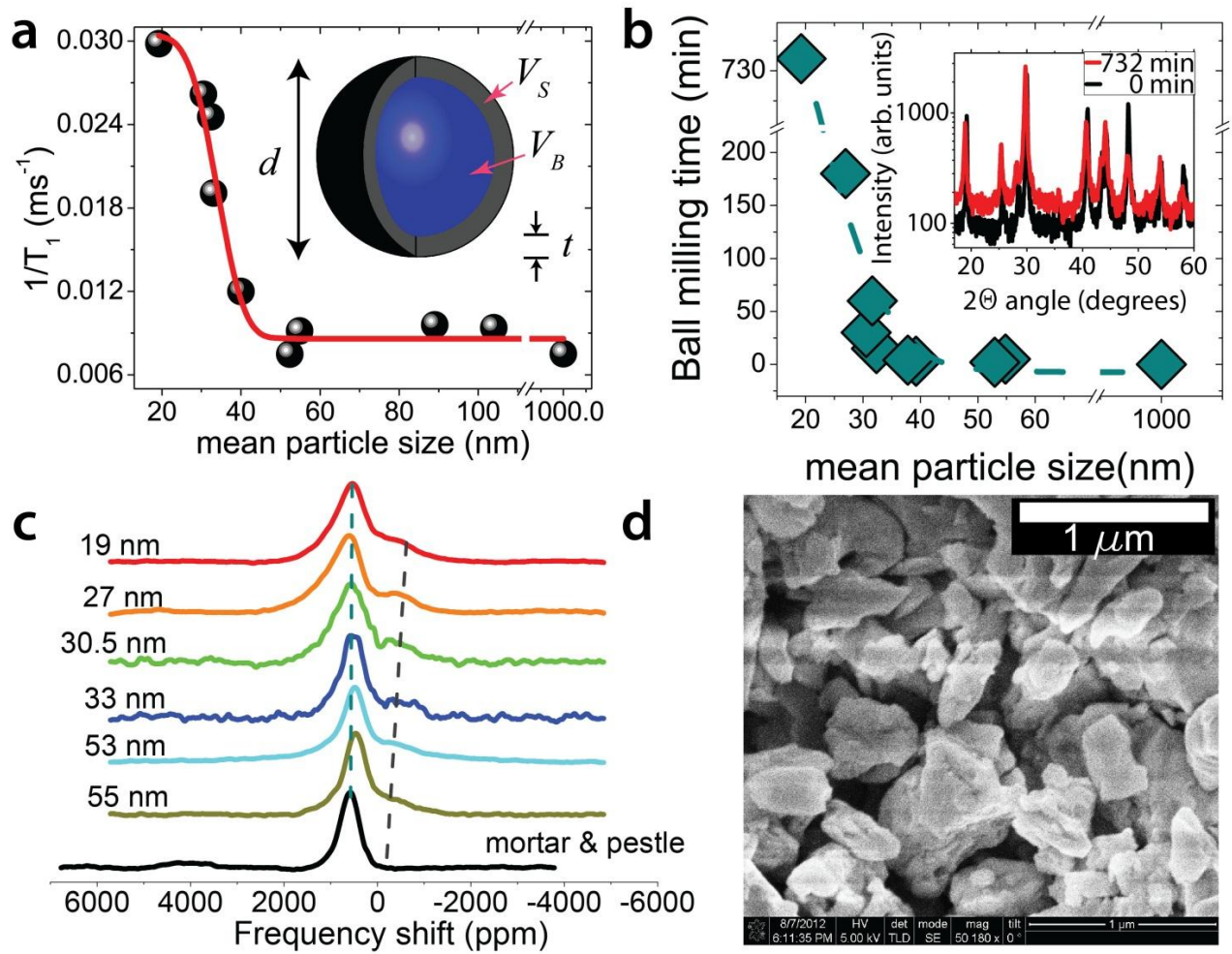

FIG. 2 (color online) $^{125}$Te spin-lattice relaxation rate as a function of the mean particle size of ball-milled $Bi_2Te_3$ nanoparticles. The mean particle size was estimated from PXRD data. The inset schematically describes the diameter [$d$], the particle core volume [bulk, $V_B$], the volume of its shell [surface, $V_S$] as well as the shell thickness [$t$]. The shell is the volume where metallic surface states reside (a). The mean particle size of the nanoparticles per each milling time process estimated from PXRD data (b). PXRD data of 19 nm (732 min) and m&p sample reveals crystallinity (b-inset). Comparison of $^{125}$Te NMR spectrum of $Bi_2Te_3$ nanoparticles versus polycrystalline $Bi_2Te_3$ (m&p sample). As the particle size decreases, a shoulder peak emerges and grows. We postulate that this shoulder peak is related to the metallic surface states (c). A representative SEM (*s*canning *e*lectron *m*icroscopy) image of particles with average size 55 nm (d).

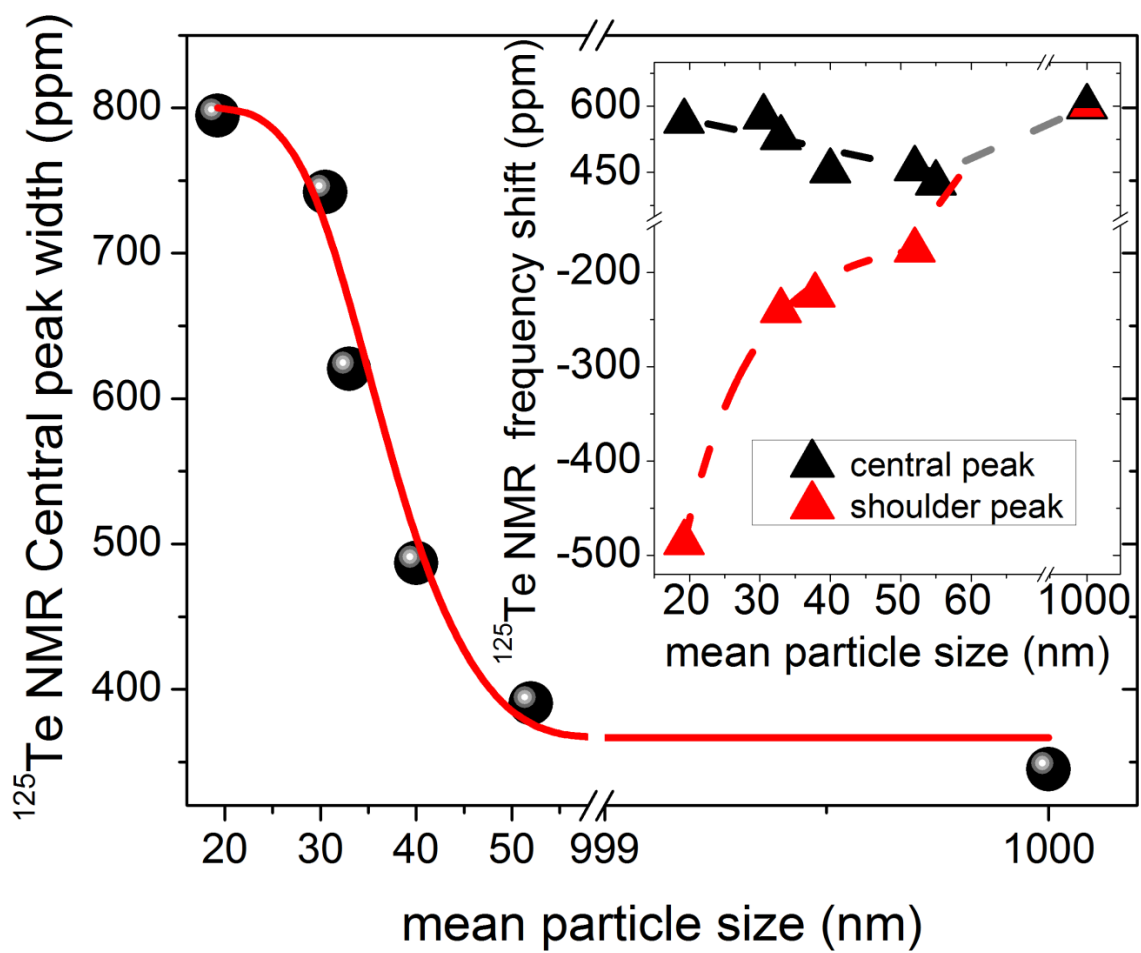

FIG. 3 (color online) $^{125}$Te NMR central peak width as function of the mean particle size of the ball-milled $Bi_2Te_3$ nanoparticles. The central and shoulder peaks are seen in Figure 2. The mean particle size was estimated from PXRD data. The inset shows the splitting of the $^{125}$Te resonance into central and shoulder peaks for nanoparticles. The splitting increases with decreasing particle size.

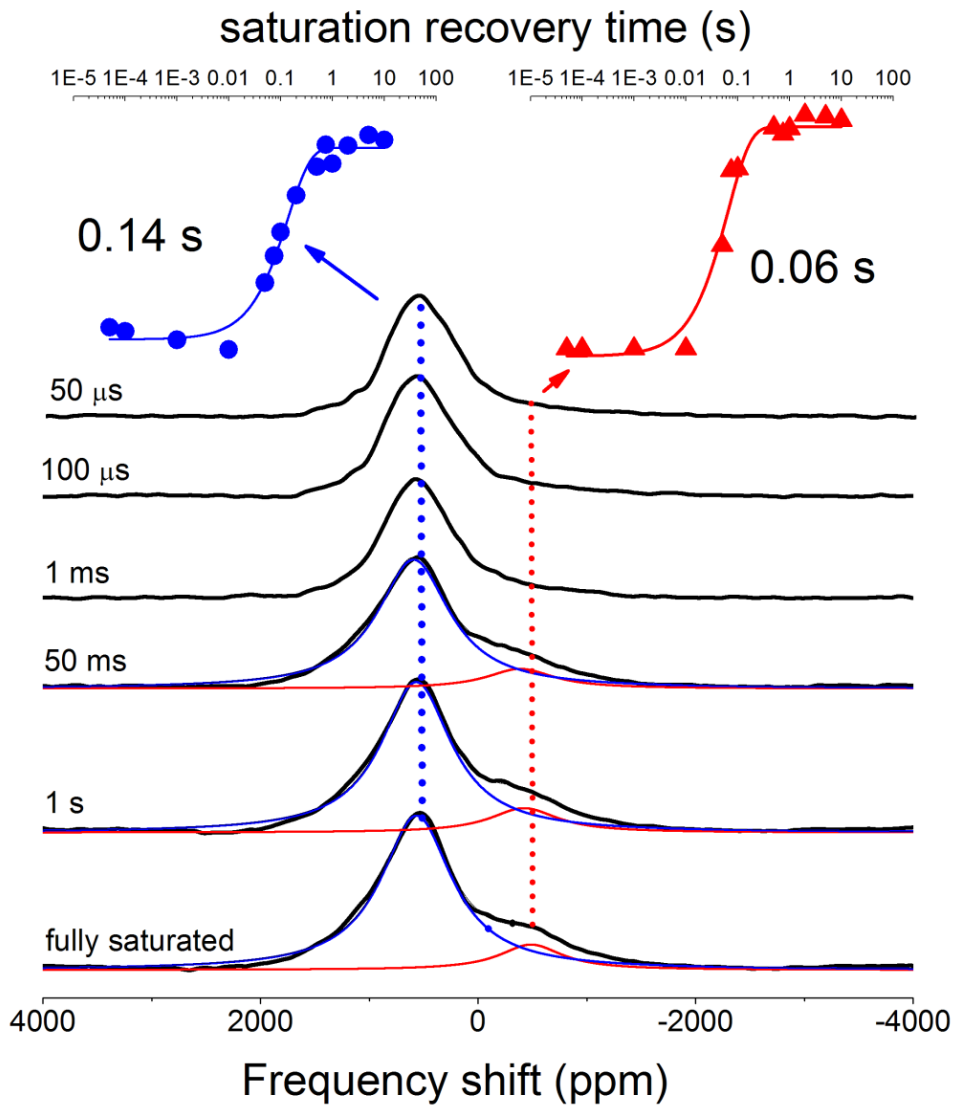

FIG. 4 (color online) $^{125}$Te NMR spectra of 19-nm ball-milled $Bi_2Te_3$ at various delay times during the saturation recovery process and the spin-lattice saturation recovery relaxation data for the "shoulder" (red line) and the central (blue line) peak, respectively. For clarity, all spectra have been rescaled to the same vertical range.

# References

[1] E. Roduner, Chem. Soc. Rev. **35**, 583 (2006).
[2] D. J. Norris, A.L. Efros and S. C. Erwin, Science **28**, 5871 (2008).
[3] D. Mocatta, G. Cohen, J. Schattner, O. Millo, E. Rabani and U. Benin, Science **332**, 77 (2011).
[4] A. Panich, M. Shao, C. L. Teske, W. Bensch, Phys. Rev. B **74**, 233305 (2006).
[5] C. B. Satterthwaite, R. Jr. W. Ure, Phys. Rev. **108**, 1164 (1957).
[6] D. A. Wright, Nature **181**, 834 (1958).
[7] S. K. Mishra, S. Satpathy, O. Jepsen, J. Phys. Condens. Matter **9**, 461 (1997).
[8] H. Zhang, C-X. Liu, X-L. Qi, X. Dai, Z. Fang, S-C. Zhang, Nat. Phys. **5**, 438 (2009).
[9] B. Yan, H-J. Zhang, C-X. Liu, X-L. Qi, T. Frauenheim, and S-C. Zhang, Phys. Rev. B **82**, 161108 (2010).
[10] M. Z. Hasan, C. L. Kane, Rev. Modern Phys. **82**, 3045 (2010).
[11] Y. Ran, Y. Zhang and A.Vishwanath, Nat. Phys. **5**, 298 (2009).
[12] R. E. Taylor, B. Leung, M. P. Lake, L.-S. Bouchard, J. Phys. Chem. C. **116,** 17300 (2012).
[13] R.K. Harris, E. D. Becker, S.M.C. De Menezes, R. Goodfellow, P. Granger, Pure Appl. Chem. **73**, 1795 (2001).
[14] D. Massiot, I. Farnan, N. Gautier, D. Trumeau, A. Trokiner, J. P. Coutures, Solid State NMR **4**, 241 (1995).
[15] T. C. Farrar, E. D. Becker, *Pulse and Fourier Transform NMR, Introduction to Theory and Methods* (Academic Press, New York, 1971).
[16] O. Kiyoshi and E. Keizo, J. Magn. Magn. Mat. **177-181**, 1443 (1971).
[17] T.Y. Hwang, I.J. Lowe, K.F. Lau, R.W. Vaughan, J. Chem. Phys. **65**, 912 (1976).
[18] D. Hilger, S.A. Kazanskii, A.I. Ryskin, W. W. Jr. Warren, Physica B **308-310**, 1020 (2001).
[19] G. Kaurr, G. Denninger, Appl. Magn. Reson. **39**, 185 (2010).
[20] W. Gou, S. Rodriguez, Y. Li, Jr. J. Ross, Phys. Rev. B **80**, 144108 (2009).
[21] A. Grykalowska, B. Nowak, Intermetallics **15**, 1479 (2007).
[22] A. Abragam, *Principles of Nuclear Magnetism*, Oxford, London (1996).
[23] L. He *et al.*, Nano Lett. **12**, 1486 (2012).
[24] D. Kong, *et al.*, Nature Nanotech. **6**, 705 (2011).
[25] D. Kong, *et al.* Nano Lett. **10**, 2245 (2010).
[26] D.P. Tunstall, S. Patou, R.S. Liu, Y.H. Kao, Materials Research Bulletin **34**, 1513 (1999).
[27] W. G. Proctor, F. C. Yu, Phys. Rev. **77**, 717 (1950).
[28] W. C. Dickenson, Phys. Rev. **77**, 736 (1950).
[29] W. D. Knight, Phys. Rev. **76**, 1259 (1949).
[30] M. F. Olekseeva, E. I. Slinko, K. D. Tovstyuk, A. G. Handozhko, Phys. Stat. Sol. (A) **21**, 759 (1974).
[31] J. Y. Leloup, B. Sapoval, G. Martinez, Phys. Rev. B **7**, 5276 (1973).
[32] S. Misran, *et al.*, J. Phys. C: Solid State Phys. **20**, 277 (1987).
[33] T.P. Das and E.H. Sondheimer, Phil. Mag. **5**, 529 (1960).
[34] A. M. Clogston, V. Jaccarino, and Y. Yafet, Phys. Rev. **134**, 3A (1964).
[35] R. E. Watson *et al.*, Phys. Rev. B **3**, 222 (1971).
[36] B.F. Williams and R.R. Hewitt, Phys. Rev. **146**, 1 (1966).
[37] E. M. Levin, B. A. Cook, K. Ahn, M. G. Kanatzidis, K. Schmidt-Rohr, Phys. Rev. B **80**,

115211 (2009).
[38] P.N. Lim *et. al*, in SIMTech technical reports (STR_V12_N3_01_FTG) of Synthesis and Processing of nanostructured thermoelectric materials  3, Vol. 12 (2011).
[39] N. Bouad, R.M. Marin-Ayral, J.C. Tedenac, J. Alloys and Compd. **297**, 312 (2000).
[40] J.  Martin, G.S.  Nolas, W. Zhang, L. Chen, Appl. Phys. Lett.  **90**, 222112 (2007).
[41] TS.  Oh, J. Choi, DB. Hyun, Scripta Metall. Mater.  **32**, 595 (1995).
[42] N. Bouad, RM. Marin-Ayral, J.S. Tedenac, J. Alloys Compd. **297**, 312 (2000).
[43] HW. Lee, DY. Lee, I.J. Kim, BC. Woo, in Proceedings ICT '02 of Thermoelectrics, Twenty- First International Conference (2002).
[44] Y.L. Chen, et. al., Science **325**, 178 (2009).
[45] D. Hsieh, Phys. Rev. Lett. **103**, 146401 (2009).